\documentclass[aps,prd,showpacs,twocolumn]{revtex4}
\usepackage{graphicx}
\usepackage{dcolumn}
\usepackage{bm}
\newcommand{\bda}{\begin{\displaymath}\begin{array}{rl}}
\newcommand{\eda}{\end{array}\end{displaymath}}
\newcommand{\be}{\begin{equation}}
\newcommand{\ee}{\end{equation}}
\newcommand{\bdm}{\begin{displaymath}}
\newcommand{\edm}{\end{displaymath}}
\newcommand{\bea}{\begin{eqnarray}}
\newcommand{\eea}{\end{eqnarray}}

\newcommand{\co}{\; \; ,}

\newcommand{\ubar}{\overline{\rule[0.42em]{0.4em}{0em}}\hspace{-0.5em}u}

\newcommand{\sbar}{\,\overline{\rule[0.42em]{0.4em}{0em}}\hspace{-0.5em}s}

\newcommand{\rvac}{\,|0\rangle}

\newcommand{\ChPT}{$\chi$PT\hspace{1mm}}
\begin{document}

\title{Mass and width of the {\boldmath $\,\sigma$}}

\author{H.~Leutwyler}

\affiliation{Institute for Theoretical Physics, University of 
  Bern\\ Sidlerstrasse 5, CH-3012 Bern, Switzerland\\
leutwyler@itp.unibe.ch}

\begin{abstract}
\begin{center}{\it Contribution to the proceedings of MESON 2006 (Krakow)}
\end{center}
I report on recent work done in collaboration with Irinel
  Caprini and Gilberto Colangelo \cite{CCL}. We observe that the Roy equations
  lead to a representation of the $\pi\pi$ scattering amplitude that
  exclusively involves observable quantities, but is valid for complex values
  of $s$. At low energies, this representation is dominated by the
  contributions from the two subtraction constants, which are known to
  remarkable precision from the low energy theo\-rems of chiral perturbation
  theory. Evaluating the remaining contributions on the basis of the available
  data, we demonstrate that the lowest resonance carries the quantum numbers
  of the vacuum and occurs in the vicinity of the threshold.  Although the
  uncertainties in the data are substantial, the pole position can be
  calculated quite accurately, because it occurs in the region where the
  amplitude is dominated by the subtractions. The calculation neatly
  illustrates the fact that the dynamics of the Goldstone bosons is governed
  by the symmetries of QCD.

\end{abstract}


\maketitle

\vspace{1em}Pions play a crucial role whenever the strong interaction is
involved at low energies -- the Standard Model prediction for the muon
magnetic moment provides a good illustration.  The present talk concerns the
remarkable theo\-re\-ti\-cal progress made in low energy pion physics in
recent years. I concentrate on work based on dispersion theory \cite{CCL}, but
also show some lattice results, obtained with quark masses that are small
enough for a controlled extrapolation to the values of physical interest to be
within reach. The low energy precision
measurements \cite{E865,DIRAC,NA48} are consistent with
the predictions and even offer a stringent test for one of these.
\begin{figure*}[thb]\vspace{-1em}
\centering
\resizebox{\textwidth}{!}{\includegraphics[angle=-90]
{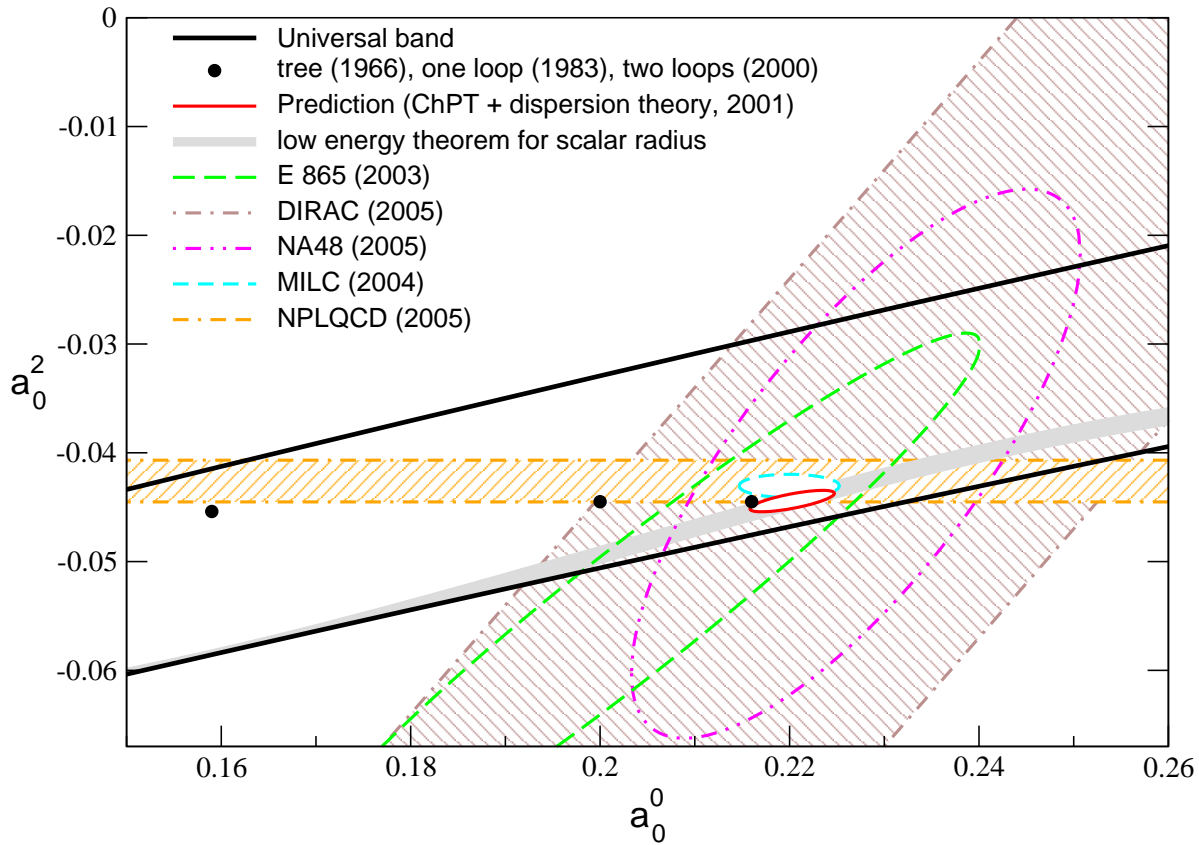}} 
\caption{\label{figa0a2}$S$ wave scattering lengths.} 
\end{figure*}

From the point of view of dispersion theory, $\pi\pi$ scattering is
particularly simple: the $s$-, $t$- and $u$-channels represent the same
physical process. As a consequence, the real part of the scattering amplitude
can be represented as a dispersion integral over the imaginary part and the
integral exclusively extends over the physical region \cite{Roy}. The
representation involves two subtraction constants, which may be identified with
the $S$-wave scattering lengths $a_0^0, a_0^2$. The projection of the
amplitude on the partial waves leads to a dispersive representation for these,
the Roy equations.

The pioneering work on the physics of the Roy equations was carried out more
than 30 years ago \cite{MP}. The main problem encountered at that time was that
the two subtraction constants occurring in these equations were not known.
These constants dominate the dispersive representation at low energies, but
since the data available at the time were consistent with a very broad range
of $S$-wave scattering lengths, the Roy equation ana\-ly\-sis was not
conclusive.

The insights gained by means of \ChPT thoroughly changed
the situation. The corrections to Weinberg's low energy theorems \cite{Weinberg
  1966} for $a_0^0, a_0^2$ (left dot in Fig.1) have been worked out to first
non-leading order \cite{GL} (middle dot) and those of next-to-next-to leading
order are also known \cite{BCEGS} (dot on the right). Very accurate predictions
for the scattering lengths are obtained by matching the chiral and dispersive
representations in the interior of the Mandelstam triangle \cite{CGL} (small
ellipse). The lattice results of the MILC collaboration \cite{MILC} also yield
an estimate for the scattering lengths.  Using their values for the coupling
constants $L_4,L_5,L_6, L_8$ of the effective chiral lagrangian and neglecting
two loop effects, we arrive at the one standard deviation contour indicated by
the small dashed ellipse. The horizontal band represents the value of the
scattering length $a_0^2$ obtained by the NPLQCD collaboration \cite{Beane}.

The plot shows that $\pi\pi$ scattering is one of the very rare hadronic
processes where theory is ahead of experiment: the two large ellipses and the
tilted band indicate the results obtained on the basis of experiments done at
Brookhaven \cite{E865} and CERN \cite{DIRAC,NA48}. Possibly, the error
estimates attached to the lattice results are too optimistic and the analysis
of the $K\rightarrow 3\pi$ data also requires further study, but it is fair to
say that all of the published results confirm the
theore\-ti\-cal predictions.

In the following, I focus on the results obtained on this basis for the low
energy properties of the isoscalar $S$-wave. The corresponding $S$-matrix
element, $S_0^0=\eta_0^0\exp 2\,i\, \delta_0^0$, is related to the partial
wave amplitude $t_0^0$ by
\be\label{S00}S_0^0(s)=1+2\,i\,\rho(s)\,t_0^0(s)\,,\hspace{2em}
\rho(s)=\sqrt{1-4M_\pi^2/s}\,.\ee The various phenomenological ana\-ly\-ses
are not in good agreement \cite{pheno}. In fact, until ten years ago, the
information about the $\sigma=f_0(600)$ was so shaky that this resonance was
banned from the data tables. The work of T\"ornqvist and Roos \cite{Tornqvist
  and Roos} resurrected it, but the estimate
$M_\sigma-\frac{i}{2}\,\Gamma_\sigma= (400\div 1200)- i\,(300\div 500)$ MeV of
the Particle Data Group \cite{PDG 2006} indicates that it is not even known for
sure whether the lowest resonance of QCD carries the quantum numbers of the
$\sigma$ or those of the $\rho$.

The positions of the poles represent universal properties of the strong
interaction which are unambiguous even if the width of the resonance turns out
to be large, but they concern the non-perturbative domain, where an analysis
in terms of the local degrees of freedom of QCD -- quarks and gluons -- is not
in sight. One of the reasons why the values for the pole position of the
$\sigma$ quoted by the Particle Data Group cover a very broad range is that
all of these rely on the extrapolation of hand made parametrizations: the
data are represented in terms of suitable functions on the real axis and the
position of the pole is determined by continuing this representation into the
complex plane. If the width of the resonance is small, the ambiguities
inherent in the choice of the parametrization do not significantly affect the
result, but the width of the $\sigma$ is not small.

We have found a method that does not invoke parametrizations of the data. It
relies on the following two observations: 

1. The $S$-matrix has a pole on the second sheet if and only if it has a zero
on the first sheet. In order to determine the pole position it thus suffices
to have a reliable representation of the scattering amplitude on the first
sheet.

2. The Roy equations hold not only on the real axis, but in a limited region
of the first sheet.  Since the pole from the $\sigma$ occurs in that region,
we do not need to invent a parametrization, but can rely on the explicit
representation of the amplitude provided by these equations.
\begin{figure}[thb]
\centering
\vspace{-1em}
\centering
\includegraphics[width=7cm,angle=-90]{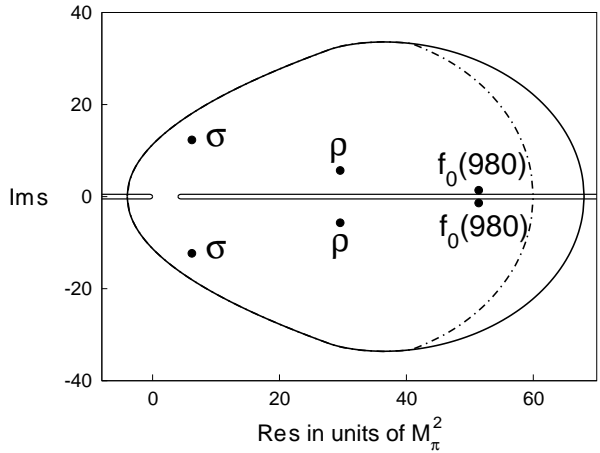}
\caption{\label{figdomain}
Domain of validity of the Roy equations.}  
\end{figure}

Roy established the validity of his equations from first principles for real
values of s in the interval $-4M_\pi^2<s< 60 M_\pi^2$. Using known results of
general quantum field theory \cite{Martin book}, we have demonstrated that
these equations also hold for complex values of $s$, in the intersection of
the relevant Lehmann-Martin ellipses \cite{CCL}. The dash-dotted curve in
Fig.\ref{figdomain} shows the domain of validity that follows from axiomatic
field theory, while the full line depicts the slightly larger domain obtained
under the assumption that the scattering amplitude obeys the Mandelstam
representation. I emphasize that the boundary does not represent a singularity
of the amplitude, but merely limits the region where the Roy equations hold in
the form given. Modified representations with a much larger domain of
validity can be found in the work of Roy and Wanders \cite{Roy Wanders 1978}
and in the references quoted therein.

For our analysis, it is essential that the dispersion integrals are dominated
by the contributions from the low energy region: because the Roy equations
involve two subtractions, the kernels fall off with the third power of the
variable of integration. The left hand cut plays an important role here: taken
by itself, the contribution from the right hand cut is sensitive to the poorly
known high energy behaviour of the imaginary parts, but taken together with
the one from the left hand cut, the high energy tails cancel. In this
connection, I note that most pole determinations assume that the left hand cut
can be neglected \cite{FN Zhou}. Since the distance between the pole and the
left hand cut is not large, this assumption is difficult to justify.

The Roy equations thus provide us with an explicit representation of the
function $S^0_0(s)$ for complex values of $s$, in terms of rapidly convergent
dispersion integrals over the imaginary parts of the partial waves. In
connection with the determination of the pole from the $\sigma$, the most
important contribution is the one from the subtraction term. The dispersion
integrals over the $S$- $P$- and $D$-waves generate a correction which can be
evaluated with available phase shift analyses -- in particular with the one
obtained by solving the Roy equations \cite{CGL}. The contributions from high
energies and high angular momenta can be estimated by means of the Regge
representation of the scattering amplitude \cite{ACGL} -- these contributions
barely affect the pole position.

For the central solution of the Roy equations, the function $S_0^0(s)$
contains two pairs of zeros in the domain of interest: \be s_\sigma = (6.2 \pm
i \,12.3)\,M_\pi^2\,,\hspace{1em}s_{\!f_0}=(51.4\pm i\, 1.4)\,M_\pi^2\,.\ee
These are indicated in Fig.\ref{figdomain}, which may also be viewed as a
picture of the second sheet -- the dots then represent poles rather than
zeros. For comparison, the figure also indicates the position of the zeros in
$S^1_1(s)$, which characterize the $\rho$.

The higher one of the two pairs of zeros represents the well-established
resonance $f_0(980)$, which sits close to the threshold of the transition
$\pi\pi\rightarrow K\bar{K}$. The corresponding pole generates a spectacular
interference phenomenon with the branch point singularity, which gives rise to
a sharp drop in the elasticity. Our analysis adds little to the detailed
knowledge of that structure.

The lower pair of zeros corresponds to a pole in the lower half of the second
sheet at \cite{CCL} \be\label{eqmsigma} M_\sigma-\frac{i}{2}\,\Gamma_\sigma
=\sqrt{s_\sigma}=441\, \rule[-0.2em]{0em}{1em}^{+16}_{-\,8}-
\,i\;272\,\rule[-0.2em]{0em}{1em}^{+\,9}_{-12.5}\;\mbox{MeV}\,.\ee The error
bars account for all sources of uncertainty. They are calculated by (a)
estimating the uncertainties in the input used when solving the Roy equations
and (b) following error propagation to determine the uncertainty in the result
for the pole position. For details, I refer to our publication \cite{CCL}.
\begin{figure}[thb]
\vspace{-1em}
\centering
\includegraphics[width=7cm,angle=-90]{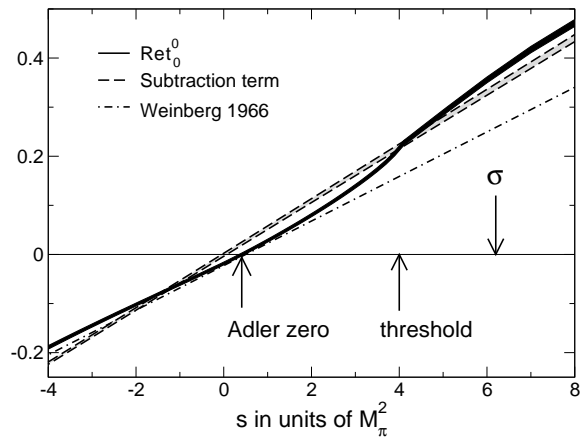}
\caption{\label{figsubthreshold}Dominance of the subtraction term.}   
\end{figure}

The reason why the pole position of the $\sigma$ can be calculated rather
accurately is that (a) the pole occurs at low energies and (b) there, the
isoscalar $S$-wave is dominated by the subtraction term. This is illustrated
in Fig.\ref{figsubthreshold}, where the real part of $t^0_0(s)$ is plotted
versus $s$. The graph demonstrates that, in the region shown, the full
amplitude closely follows the subtraction term. The final state interaction
does generate curvature -- in particular, the cusp at $s=4M_\pi^2$ is visible
-- but since Goldstone bosons of low momentum interact only weakly, the
contributions from the dispersion integrals amount to a small correction. In
the language of \ChPT$\!$, the dispersion integrals only show up at NLO.
Moreover, at leading order, the subtraction constants are determined by the
pion decay constant. Dropping the dispersion integrals and inserting the
lowest order predictions for the scattering lengths, the Roy equation for the
isoscalar $S$-wave reduces to the well-known formula, which Weinberg derived
40 years ago \cite{Weinberg 1966}, \be\label{eqWeinberg} t_0^0(s)
=\frac{2\,s-M_\pi^2}{32\,\pi\, F_\pi^2}\co\ee and which is shown as a
dash-dotted line in Fig.\ref{figsubthreshold}.

The main feature at low energies is the occurrence of an Adler zero,
$t^0_0(s_A)=0$. At LO, the zero is at $s_A=\frac{1}{2}M_\pi^2$. The higher
order corrections generate a small shift, which can be evaluated from our
solution of the Roy equations. The uncertainties in the result, $s_A=(0.41\pm
0.06)\,M_\pi^2$, are dominated by those in our predictions for the scattering
lengths. While the Adler zero sits below the threshold, at a real value of
$s$, the lowest zeros of the $S$-matrix, $S^0_0(s_\sigma)=0$, occur in the
region $\mbox{Re}\hspace{0.05em}s_\sigma >4 M_\pi^2$, with an imaginary part
that is about twice as large as the real part.

The formula (\ref{eqWeinberg}) explains why the $S$-matrix has a zero
in the vicinity of the threshold. In this approximation, the zero of
$S_0^0(s)$ occurs at $\sqrt{s}= 365 -i\, 291$ MeV. The number differs from the
``exact'' result in Eq.~(\ref{eqmsigma}) by about 20 percent. The dispersion
integrals are essential for the partial wave to obey unitarity, but only
represent a correction.  In view of this, the precision in
Eq.~(\ref{eqmsigma}) is rather modest.

I conclude that the same theoretical framework that leads to incredibly sharp
predictions for the threshold parameters of $\pi\pi$ scattering \cite{CGL} also
shows that the lowest resonance of QCD carries the quantum numbers of the
vacuum.  The physics of the $\sigma$ is governed by the dynamics of the
Goldstone bosons: the properties of the interaction among two pions are
relevant \cite{Markushin and Locher}. In quark model language, the wave
function contains an important tetra-quark component \cite{Jaffe}.

The resonance $f_0(980)$ is also governed by Goldstone boson dynamics -- two
kaons in that case. Very recently, the method described above was applied to
the case of $\pi K$ scattering \cite{Descotes-Genon and Moussallam 2006}. In
this case, the analogue of the back-of-the-envelope calculation sketched above
relies on the tree level approximation for the $I=\frac{1}{2}$ $S$-wave
obtained from the effective SU(3)$\times$SU(3) lagrangian and yields $m_\kappa
= 671-i\,262$ MeV, remarkably close to the ``exact'' value, obtained from the
solution of the Roy-Steiner equations, $m_\kappa =(658\pm 13)-i\,(278.5\pm
12)$ MeV.  Evidently, the physics of the $\kappa$ is very similar to the one
of the $\sigma$.

The BES data \cite{BES sigma} on the decay $J/\psi\rightarrow\omega\pi\pi$ play
a prominent role in scalar meson spectroscopy. Although I did not discuss this
at the conference, I add a comment concerning the recent claim \cite{Bugg sigma
  pole} that our analysis is in conflict with these data. I denote the
projection of the decay amplitude onto the $\pi\pi$ configuration with
$I=\ell=0$ by $A_0(s)$.  If rescattering on the $\omega$ and the inelasticity
due to $4\pi$ final states is neglected, unitarity implies that $A_0(s)$
shares an important property with the scalar pion form factor
$F_0(s)=\langle\pi^+\pi^-\,\mbox{out}|\,\ubar u\rvac$: below the $K\bar{K}$
threshold, the phase coincides with the phase shift $\delta_0^0(s)$ of
$\pi\pi$ scattering \cite{Caprini}. On the interval $4M_\pi^2< s < 4 M_K^2$,
the ratio $R(s)=A_0(s)/F_0(s)$ must therefore be real and can vary only
slowly.

\begin{figure*}[thb]
\vspace{-1em}
\includegraphics[width=6.2cm,angle=-90]{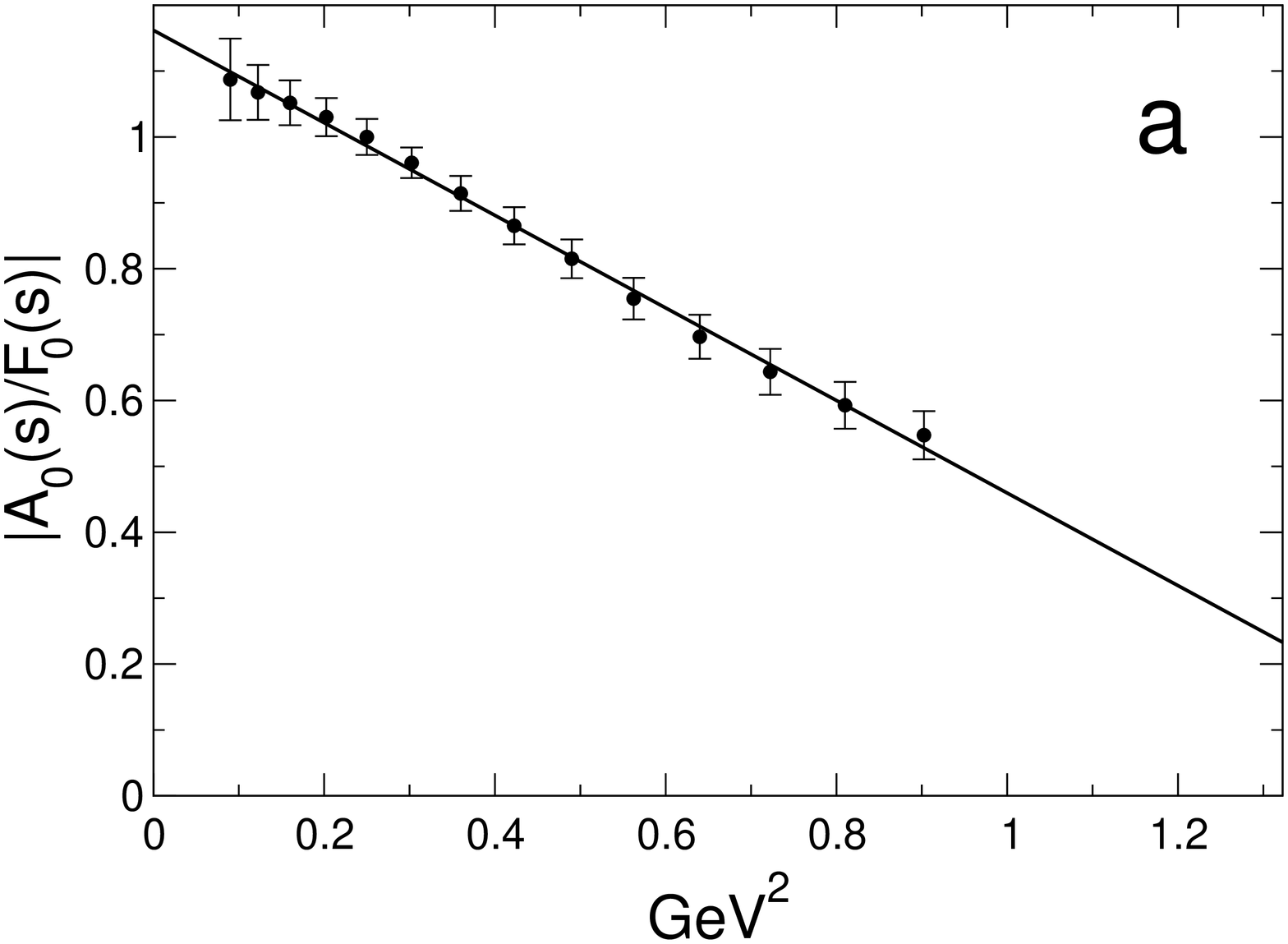}
\includegraphics[width=6.2cm,angle=-90]{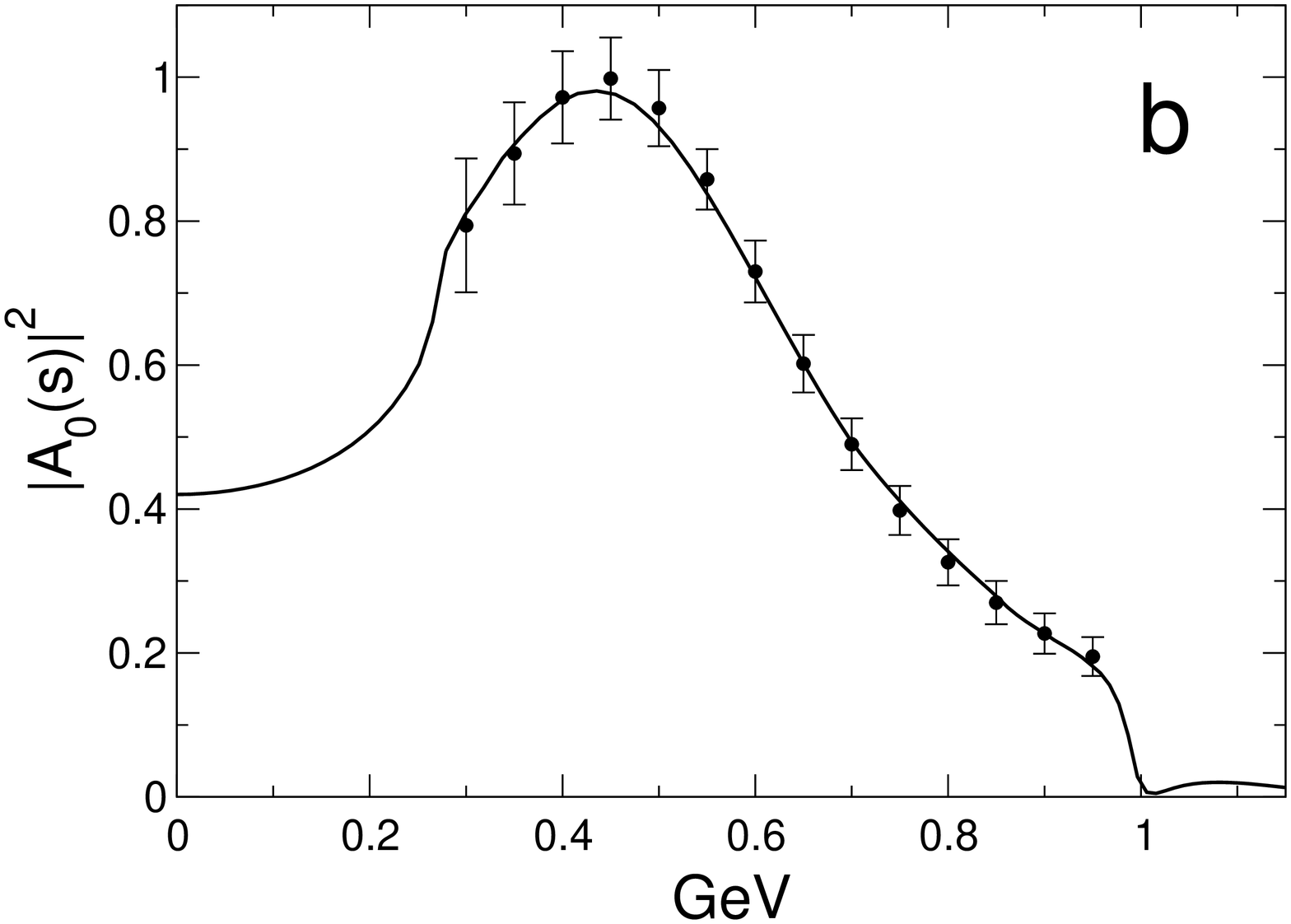}
\caption{\label{figBES}$J\psi\rightarrow\omega\pi\pi$ decay: energy dependence
  of the amplitude $|A_0(s)|$.}  
\end{figure*}

Fig.\ref{figBES}a shows the magnitude of this ratio, obtained by dividing the
outcome of the partial wave analysis \cite{FN Bugg} for $|A_0(s)|$ with the
dispersive result for the scalar form factor \cite{ACCGL} (arbitrary units on
the vertical axis). The ratio indeed varies only slowly with the
energy \cite{FN Lahde}. 
This demonstrates that the profile of the BES data on $|A_0(s)|$ closely
resembles the one of the scalar form factor. In either case, there is a peak
around $\sqrt{s}\simeq M_\sigma$, but the phase reaches $90^\circ$ only when
the energy becomes about twice as large as the mass of the $\sigma$. 

The low energy behaviour of the various amplitudes can be understood on the
basis of twice subtracted dispersion relations.  In the case of $\pi \pi$
scattering, the subtraction constants are predicted by chiral symmetry.  As a
consequence, the low energy behaviour of the phase shift $\delta^0_0(s)$ and
the position of the pole from the $\sigma$ are understood. The symmetry, in
particular, requires the occurrence of an Adler zero, which suppresses the
scattering amplitude at low energies, so that the peak in that amplitude is
pushed up, to $\sqrt{s}\simeq 2\,M_\sigma$. There is no such suppression in
the decay amplitude, nor is there a prediction in that case: the two
subtraction constants in the dispersion relation for
$R(s)=A_0(s)/F_0(s)$,\be\label{eqR}
R(s)=R_0+R_1\,s+\frac{(s-2M_K^2)^2}{\pi}\!\int\!
\frac{dx\,\mbox{Im}R(x)}{(x-2M_K^2)\hspace{-0.05em}\rule{0em}{0.6em}^2\,
  (x-s)}\co\ee must be taken from experiment. The integral receives a
contribution from $s>4M_K^2$, as well as one from $s<0$. For definiteness, I
have identified the subtraction point with the center of the interval between
these two cuts. The straight line shown in Fig.\ref{figBES}a is obtained by
setting $s_0\equiv -R_0/R_1=1.65\,\mbox{GeV}^2$ and simply dropping the
dispersion integral.  Fig.\ref{figBES}b shows that the approximation
$A_0(s)\simeq R_1(s-s_0)F_0(s)$ indeed accounts for the pronounced energy
dependence seen in the BES data.

Above the range covered by the data points, the dispersive contributions
presumably become quite important. In particular, the behaviour near the
$K\bar{K}$ threshold need not be the same as the one of the non-strange form
factor $F_0(s)$. As can be seen in Fig.\ref{figBES}a, this form factor has a
narrow dip there. The form factor of the operator $\sbar s$ instead exhibits a
narrow peak \cite{ACCGL}. If all final states other than $\pi\pi$ and
$K\bar{K}$ are ignored, the general solution of the unitarity conditions is a
linear combination of the two. An experimental investigation of the structure
of $A_0(s)$ in the region of the $f_0(980)$ could shed some light on the
importance of strange quarks in this context. At any rate, the zero, which a
linear approximation for the ratio $R(s)$ necessarily entails, sits far beyond
the region where that approximation is meaningful.

I do not see any reason to doubt that the BES data and the partial wave
analysis thereof are correct. These clearly reveal the presence of the
$\sigma$, but the pole positions extracted from simple Breit-Wigner
treatments \cite{BES sigma} of the amplitude $A_0(s)$ or more elaborate
models \cite{Bugg sigma pole} are subject to large theoretical uncertainties:
the extrapolation off the real axis is sensitive to details of the
parametrization that are not understood. In the language of dispersion theory,
the problems arise from (a) the inelastic channels $\pi\pi\rightarrow
K\bar{K}$ and $\pi\pi\rightarrow\eta\eta$, (b) the $f_0(980)$, (c) the left
hand cut and (d) rescattering on the $\omega$.  Unitarity and crossing
symmetry very strongly constrain the scattering amplitude, but barely tell us
anything about the transition $J/\psi\rightarrow\omega\pi\pi$. 

We have applied our method to the models for the scattering amplitude proposed
by Bugg \cite{Bugg sigma pole}. The outcome for the mass and width of the
$\sigma$ is close to the central values in Eq.~(\ref{eqmsigma}). I conclude
that our pole position is consistent with the BES data. There is a conflict
only with the values for the mass and width obtained from the extrapolation of
those models to complex values of $s$. It arises because the quoted errors do
not account for the uncertainties inherent in such
extrapolations \cite{Pennington}.


\begin{thebibliography}{0}
\bibitem{FN Lahde}The model for $A_0(s)$ of L\"{a}hde and
  Mei{\ss}ner \cite{Lahde and Meissner} corresponds to $R(s) = \mbox{const}.$
  The form factor is taken from a unitarized version of \ChPT$\!\!$. It is
  interesting to note that, in the version proposed by Hannah \cite{Hannah},
  the unitarized form factor contains a second sheet pole at $445-i\,235$ MeV.
\end{thebibliography}
\end{document}